\documentclass[aps,reprint,prb,twocolumn,amsmath,amssymb]{revtex4}
\usepackage{fancyhdr}
\usepackage{graphicx}
\usepackage{graphicx} 
\usepackage{dcolumn}  
\usepackage{bm}       
\usepackage{amssymb}
\usepackage{amsmath}
\usepackage{color}
\usepackage{psfrag}

\def\bea{\begin{eqnarray}}
\def\eea{\end{eqnarray}}
\def\ben{\begin{equation}}
\def\een{\end{equation}}
\def\benu{\begin{enumerate}}
\def\enu{\end{enumerate}}
\def\bit{\begin{itemize}}
\def\eit{\end{itemize}}

\def\n{n}

\def\sss{\scriptscriptstyle\rm}

\def\g{_\gamma}

\def\half{\frac{1}{2}}

\def\br{{\bf r}}

\def\s{_{\sss S}}

\def\N{_{\sss N}}

\def\loc{^{\rm loc}}

\def\GEA{^{\rm GEA}}

\def\TF{^{\rm TF}}

\def\unif{^{\rm unif}}

\def\max{_{\rm max}}

\def\bay{\begin{array}}
\def\eay{\end{array}}
\def\beit{\begin{itemize}}
\def\eit{\end{itemize}}

\def\e{_\text{e}}

\def\e{\epsilon}

\def\m{_{\sss m}}
\def\F{_{\sss F}}
\def\loc{^{\mathrm{loc}}}

\def\WKB{^{\rm WKB}}
\def\d{\mathrm{d}}

\def\sec#1{\section{#1}}
\def\ssec#1{\subsection{#1}}

\def\crap#1{The following is crap:\\ #1}
\def\crap#1{\bf !!!!Crap was removed here!!!!}
\def\bei{\begin{itemize}}
\def\eei{\end{itemize}}
\def\benu{\begin{enumerate}}
\def\enu{\end{enumerate}}

\def\sc{_{\rm sc}}
\def\s{_{\rm s}}
\def\osc{_{\rm osc}}
\def\O{{\cal O}}
\def\N{\tilde{N}}

\begin{document}







\title{Leading corrections to local approximations}
\author{Attila Cangi}
\author{Donghyung Lee}
\author{Peter Elliott}
\author{Kieron Burke}

\affiliation{Departments of Chemistry and of Physics,
University of California, Irvine, CA 92697, USA}

\date{\today}

\begin{abstract} 

For the kinetic energy of 1d model finite systems 
the leading corrections to local approximations as 
a functional of the potential are derived 
using semiclassical methods.  The corrections are simple, 
non-local functionals of the \emph{potential}. 
Turning points produce quantum oscillations leading to 
energy corrections, which are completely different from the
gradient corrections that occur in bulk systems with slowly-varying densities.
Approximations that include quantum corrections are typically 
much more accurate than their local analogs.  The consequences
for density functional theory are discussed.

\end{abstract}


\maketitle





\sec{Introduction}

Modern density functional theory (DFT) has become a popular
electronic structure method, because of its balance between
computational efficiency and accuracy.
The original density functional theory was that of Thomas\cite{T27}
and Fermi\cite{F28} (TF), in which all parts of the electronic Hamiltonian
are approximated by explicit density functionals, and the
energy minimized over possible densities.  Typical
errors in TF theory are of order 10\% of the total energy, but
molecules don't bind.\cite{T62}  In the 1950's, Slater\cite{S51}
introduced the idea of {\em orbitals} in DFT, i.e., solving
a set of single-particle equations to construct the energy
of the interacting system, which is typically much more accurate.
This was shown to be a formally exact approach in the celebrated
works of Hohenberg and Kohn\cite{HK64} and Kohn and Sham.\cite{KS65}
The latter also introduced the local density approximation (LDA)
for the only unknown needed to solve the Kohn-Sham (KS) equations,
the exchange-correlation (XC) energy as a functional of the density.  
Since then, the field has gradually
evolved with improvements in computational power, algorithms, and
approximate functionals to the workhorse it is today.\cite{M04}

Unfortunately, the existence theorems give no hint of how to construct
approximate functionals.
Presently, there is a dazzling number of such approximate functionals
suggested in the literature, and implemented in standard codes,
both in physics and chemistry.\cite{RCFB09}
Many of these are physically 
motivated, and work well for the systems and properties for which
they were designed, but usually fail elsewhere.  There appears
to be no systematic approach to the construction of these functionals,
beyond artful constraint satisfaction.\cite{PRCS09}

In the present paper, we return to the origins of 
DFT and ask, what are the leading corrections to the original
approximation of Kohn and Sham, the LDA?
This is a very difficult question that we can only aspire to answer for
the XC energy for any electronic system.
In the present article, we answer
the question for an extremely simple case, but one that contains
many features relevant to the problems of electronic structure.

The original works on density functional approximations emphasize the
gradient expansion,\cite{HK64,KS65} which is a particular approach to 
improve upon a local density approximation. Imagine an infinitely extended 
slowly-varying gas.  The corrections to the local approximation are given
accurately by leading corrections in the density gradient. 
But real systems do not look like slowly-varying gases.  All finite
systems have evanescent regions, as do many bulk solids.
The regions can be separated via classical turning-point surfaces,
evaluated at the Fermi energy of the system.\cite{KM98}  Typically, such regions
are atomic sized.  Most importantly, the gradient expansion fails
completely both near and outside these surfaces. 
Generalized gradient approximations (GGAs) 
and other methods have been developed to overcome these difficulties.
These include only a finite order of gradients, but employ a form
which contains many powers those gradients.

To study the effects of confinement to finite regions on density 
functional approximations, we
use non-interacting particles, and study only their kinetic energy,
which was locally approximated in the original TF theory.
We study only one dimension, where semiclassical approximations are 
simple, and the WKB\cite{W26,K26,B26} form applies in the absence 
of classical turning points where the potential $v(x)$ has finite slope.  
We avoid such turning points 
by using box boundary conditions and studying only systems 
whose chemical potential is everywhere above $v(x)$.

The answer is surprising:  For most systems, the leading corrections
(in a sense that will become clear within)
are {\em not} the simple gradient corrections commonly discussed,
and used as starting points to construct GGAs.
Instead, both the density and kinetic energy density 
can be very accurately approximated as functionals of the {\em potential}.

The limit we discuss is also an important result in itself.  We carefully show
precisely how TF becomes exact in a semiclassical limit.  Essentially,
we take $\hbar \to 0$, keeping the chemical potential $\mu$ roughly fixed, and
renormalizing the density so as to retain the original particle number.
If, further, one performs a moving-average over the space coordinate,
with a range chosen to be small compared to the spatial variation of
the potential, but large compared to quantum oscillations as $\hbar\to 0$,
the density uniformly converges to that of TF theory.
We call this the continuum limit of a {\em finite} system.
The separation between quantum eigenvalues becomes infinitesimal, and
all sums become integrals.  The integrands within contain purely classical
quantities in terms of the potential, $v(x)$.  A similar simplification
occurs for the kinetic energy density, given in terms of $v(x)$, and when
$v(x)$ is eliminated from the two expressions, what remains is the
LDA to the kinetic energy.

Having carefully defined this limit, we can then discuss the approach
to that limit, and the leading corrections to the local approximation.
We find that the dominant corrections  (in $1/\hbar$) are {\em not} gradient corrections
due to the variations in $v(x)$ in the interior, but rather are
quantum oscillations due to the hard walls at the boundaries.  
These quantum oscillations are generic features of any quantum system, and their
nature is determined by the classical turning points.\cite{KSb65}  They give rise to
the phase corrections to the classical density of states in the Gutzwiller
trace formula,\cite{G71} and will be present for {\em any} finite quantum system.
The only case in which they vanish is that of periodic boundary conditions
with the chemical potential above the maximum of the potential.  Only in such systems
does the gradient expansion produce the correct asymptotic expansion in powers
of $\hbar$, equivalent to gradients of the potential. For any finite system,
the series eventually diverges, but truncation at a lower order can yield 
highly accurate results, if the gradients are sufficiently small.

\begin{figure}[htb]
\begin{center}
\includegraphics[angle=0,height=6cm]{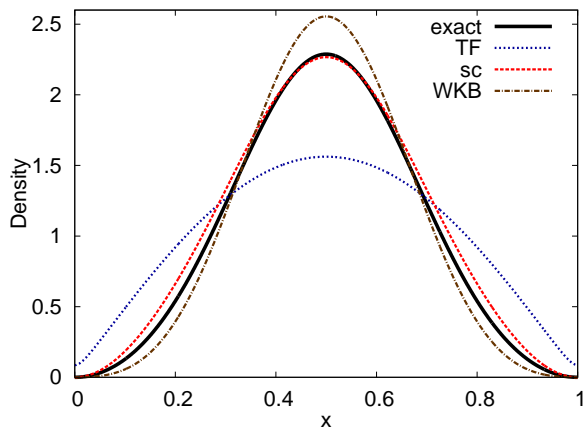}
\end{center}
\caption[Density and approximations]
{Exact, TF, our semiclassical (sc), and WKB
density for a single particle in $v(x)=-12\sin^2(\pi x)$ 
with hard walls at $x=0$ and $x=1$ ($\hbar=m=1$).}
\label{f:illus}
\end{figure}
To give some idea of the power of the methods we develop, 
in Fig.~\ref{f:illus} we show the density of one particle in a simple well 
($v(x)=-D\sin^2{(\pi x/L)}$ in a box from $0$ to $L$, with $D=12$.)  
The exact density is found by numerically solving the Schr\"odinger equation.
The TF density is found by minimizing the 1d TF kinetic energy functional
and choosing the chemical potential to yield one particle.  That density
smoothly follows the potential. Due to the hard walls 
there are no classical turning points where $v(x)$ has finite slope.
Hence, a WKB treatment can be applied here,
yielding a WKB eigenvalue that is positive and reasonably accurate.  But the
result of our present analysis is a simple formula for the density, which
uses WKB wavefunctions as input, is much more accurate still.
Perhaps more importantly, we have been unable to create situations where
our approximation completely fails.

In the electronic structure problem,
the local approximation to the XC energy is analogous 
to the TF approximation in Fig. \ref{f:illus}.  While there are many
excellent approximations that improve over the local approximation
by typically an order of magnitude, they are usually tailored to specific
systems or properties, and contain either empirical parameters\cite{Bb93} or at least
careful selection of exact conditions to impose on an approximation to
fix the parameters non-empirically.\cite{PBE96}
Our formulas are derived via a semiclassical analysis that yields unique
approximations, are robust, and typically {\em two} orders of magnitude 
better than the local approximation.

The semiclassical result for the density was given in a short report\cite{ELCB08}, but the 
kinetic energy density formula derived there failed close to the boundaries.
Here we explain that failure, in terms of boundary-layer theory\cite{BO99,N73}, but applied
to sums over eigenstates rather than individual solutions to a differential
equation.  By identifying our limit, we can then cleanly separate the two
different length scales in the problem.  
Earlier approaches\cite{KSb65,G66}
yield only the asymptotic result in the interior of the system (ironically
labelled the outer region in boundary layer theory), but that there exists a
region close to each wall (appropriately called a boundary layer) where that
solution fails.  However, within the boundary layer, a different asymptotic expansion
applies and, by matching these two solutions, we construct a {\em uniform}
approximation that is asymptotic to a given order {\em everywhere} in the
system.  
Many different aspects of these issues have been addressed over the decades
since the original work of Kohn and Sham.\cite{KSb65}  For example, Balian
and Bloch,\cite{BB71} in the context of nuclear physics, identified the
need for spatial averaging to approach the limit.
In the early 1970's, Yuan and Light\cite{LY73,YLL74}  
and coworkers\cite{LL75} developed the theory in terms
of path integrals and density matrices.
Recently, Ullmo, Baranger, and coworkers\cite{UNTB01,UJYB04} studied the 
nature of quantum oscillations in application to quantum dots.

What is the significance of our results for the real world of 3d, interacting
electrons?  Our results, for a very different case, reveal the nature of the 
corrections to local approximations. These will differ in detail depending 
on the dimensionality of the system, or the specific functional being approximated.
However, qualitative features (such as the local approximation becoming exact in 
the classical continuum limit, gradient expansions being invalid near turning
points, etc.) are general. Thus our analysis can (and already has \cite{PRCV08})
provided guidance for the construction of new XC density functionals.
On the other hand, there is also a considerable amount of work done 
in the field of orbital-free DFT,\cite{WCW96,LC05} but effort is focused 
on finding an accurate approximation to the non-interacting kinetic energy 
as an explicit functional of the density.
The present work derives potential functional approximations, an entirely 
different matter, and so has no overlap with existing work in that field.
Our work suggests that the potential is a better variable than the density,
and we show how corrections to local approximations of the density and
kinetic energy density can be derived as potential functionals 
for simple model systems, but the methods and results shown here do not
readily generalize to three dimensions.

This paper is divided as follows:
In Sec.~\ref{s:clascont}, we introduce our notation 
and define the continuum limit and show that local approximations
become exact in this limit.  Next, in Sec.~\ref{s:leadcor} 
we derive the leading corrections for both
the density and kinetic energy density as functionals of the potential
by explicit summation of WKB orbitals.  Then, in Sec.~\ref{sec:props} 
we ``fix" the difficulties at the walls to produce a uniform approximation
everywhere, and then study its properties comparing to the exact result
both in the classical continuum and the large-$N$ limits.
Finally, in Sec.~\ref{s:conseq} we discuss the implications 
and relevance for real electronic structure calculations.
In the appendix we give a detailed derivation of the interior solution of
the density and the KED using the WKB Green's function in the complex plane, 
just as has been done before.

\sec{Classical continuum}
\label{s:clascont}

In this section, we introduce our notation and briefly review the
salient points known from the literature.
As discussed in the introduction, we restrict ourselves to
non-interacting particles in one dimension.  
We use atomic units throughout ($e^2=\hbar=m_e=1$), so that
lengths are expressed in Bohr radii, and energies in hartree.

\ssec{Background and notation}

We write the Hamiltonian as
\ben
\hat h = \hat t + \hat v = -\half \frac{d^2}{dx^2} + v(x),
\een
where a hat denotes an operator with $\hat t$ being the
kinetic energy operator, and  $v(x)$ the potential.
We denote the solutions to the 
Schr\"odinger equation as
\ben
{\hat h}\, \phi_j (x)  = \epsilon_j\, \phi_j (x),~~~~~j=1,2,...\ .
\een
The solutions to the Schr\"odinger equation can be expanded
in powers of $\hbar$,\cite{Gb05} and retaining 
just the first two, we find the WKB solutions
for a given energy $\e$ are
\ben
\phi\WKB(x) = \frac{1}{\sqrt{k(x)}}\, e^{\, i\, \theta(x)}\ ,
\label{WKB}
\een
and its complex conjugate,
where the dependence on $\e$ is via the definitions of the wavevector
\ben
k(x) = {\sqrt{2 (\e-v(x))}}
\een
and classical phase 
\ben
\theta(x) = \int^x dx'\, k(x')\,,
\een
where the constant is arbitrary.
These solutions are exact when the potential is constant, and highly
accurate when the potential is slowly varying on a scale determined
by the energy.   However, the particular combination that forms an 
eigenstate depends on the boundary conditions. The density is then
\ben\label{nTFapp}
\n(x) \approx \frac{1}{\pi}\int_{-\infty}^{\mu}d\e\,|\phi\WKB(\e,x)|^2 
= \frac{k_{\mu}(x)}{\pi},
\een
where $k_{\mu}(x)$ is the wavevector evaluated at $\mu$, the chemical potential
for the system, found via normalization.
Note that $k_{\mu}(x)$ is a function of $\mu-v(x)$ alone, so we define the
{\em local} chemical potential:
\ben
\mu(x) = \mu-v(x).
\een
Then, because WKB is exact for an infinitely extended system 
with constant potential (free particles), we find
\ben
\n\unif(\mu) = \frac{1}{\pi}(2\mu)^{1/2},~~~~
t\unif(\mu)= \frac{1}{6\pi}(2\mu)^{3/2}.
\een
The corresponding integrals are local {\em potential} approximations (LPAs) 
to the particle number and kinetic energy:
\bea
N\loc[\mu(x)]&=&\int dx\, n\unif(\mu(x)),\nonumber\\
T\loc[\mu(x)]&=&\int dx\, t\unif(\mu(x)).
\eea
Inverting the relation $\n\unif[\mu(x)]$ and inserting into $t\unif[\mu(x)]$
yields the local {\em density} approximation to $T$:
\bea\label{Tlda}
T\loc[\n] =\int dx\, t\unif(\n(x)),~~~
t\unif(\n(x)) = \frac{\pi^2\n^3(x)}{6}.
\eea
This is the one-dimensional analog of the TF kinetic energy density functional
(up to simple factors of 2 for double occupation).\cite{SPb99}

One can also work backwards from the LDA to the WKB results.
The LDA for the kinetic energy allows us to find an approximate
density for a given potential, by minimizing the total energy
subject to the constraint of a given particle number.  This produces
\ben
\frac{\pi^2}{2} \n^2(x) + v(x) = \mu\,,
\een
the TF equation for this problem, which is identical to Eq.~(\ref{nTFapp}).
The solution is the TF density, $n\TF(x)=\n\unif(\mu(x))$.
The total particle number $N$ is a continuous monotonic function
of the parameter $\mu$ that is invertible for $\mu > v_{min}$, the minimum of the potential.

The leading corrections to the WKB wavefunctions, i.e., the next two
powers in $\hbar$, are well-known\cite{BO99} and  produce constant
corrections to both the phase and the wavevector.
Samaj and Percus\cite{SPb99}  
showed very elegantly how the series can be generated to any desired order.
Continuing with the higher-order corrections to WKB leads to corrections
that depend on derivatives of the potential.  
where $v'(x)=dv/dx$.
The potential gradient expansion for the density is
\ben
\n\GEA[\mu(x)](x) = \n\unif(\mu(x))
\left(1- \frac{v''(x)}{12 k_{\mu}^4(x)} + \frac{v'^2(x)}{8 k_{\mu}^6(x)} +\dots\right)\,, \label{ngea}
\een
and for the kinetic energy density 
\ben
t\GEA[\mu(x)](x) = t\unif(\mu(x))
\left(1- \frac{3v''(x)}{4k_{\mu}^4} + \frac{5 v'^2(x)}{8k_{\mu}^6(x)} +\dots\right)\,, \label{tgea}
\een
which, when inverted leads to the density {\em gradient expansion} for $T$:
\ben\label{Tgea}
T[\n] \approx \frac{\pi^2}{6} \int dx\, \n^3(x)
- \frac{1}{12} \int dx \left[\frac{\n'^2(x)}{2\n(x)}+ n''(x)\right] +\dots \ .
\een
A gradient expansion approximation is the finite truncation of that series.
Because the semiclassical
expansion is asymptotic, this is an asymptotic expansion
for periodic systems where $\mu$ is above the maximum of the potential.
It can be made arbitrarily accurate by application to sufficiently smooth
densities, but for any given density, addition of sufficient terms will
eventually lead to divergence.  For example, a potential that contains steps
will produce divergences beyond the lowest order. Also, Coulomb potentials are
known to vary too rapidly for such expansions to apply.\cite{L37}
We note that, in 1d, because of the negative coefficient in the gradient
correction, minimizing the total energy is unbounded and nonsensical in 
the presence of this correction.\cite{HKM91}  We also note that $\hbar$ never appears 
in the functional dependence on the density in Eq. (\ref{Tgea}).

\ssec{The classical continuum limit}

We define a continuum as any region of energy in which the eigenvalues of the
Hamiltonian are not discrete.  The first, simplest example is that of a particle
in a well, with a potential set that vanishes as $|x|\to\infty$.  For $\e > 0$, there is
the free-particle continuum, with scattering states of the system
that cannot be box-normalized.
Another continuum arises in solid-state physics, when we apply periodic boundary
conditions to our potentials, in order to simulate bulk matter.  Then, for
single-particle states, the energy levels form distinct bands, usually labelled
by a wavevector.  Within each band, the energy is continuous.  We call this
the bulk or thermodynamic continuum.

But any system also has a classical continuum, which can be found by letting $\gamma \to 0$,
where we have replaced $\hbar$ by $\gamma\hbar$.
As $\gamma$ becomes very small, the discrete levels of the system merge, and the envelope
of their density of states approaches a well-defined limit.  We call this limit the
classical continuum.  While it has been long understood that local density approximations
become exact in this limit,\cite{DG90} relatively little attention has been paid to 
how exactly this limit is reached in a quantum system.

\begin{figure}[htb]
\begin{center}
\includegraphics[angle=0,height=6cm]{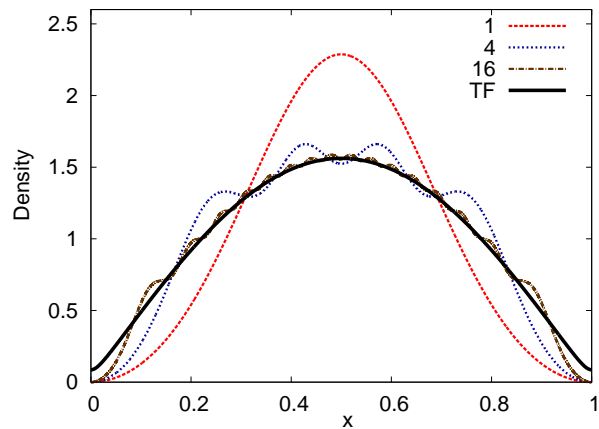}
\end{center}
\caption
{TF and renormalized exact densities for $N'=1,\, 4,\, \rm{and}\, 16$ particles
in $v(x)=-12\sin^2(\pi x),\, 0 \leq x \leq 1$, showing approach to continuum limit.}
\label{f:TFlimit}
\end{figure}

Consider a 1d box of length $L$ with given potential $v(x)$ and
particle number $N$, i.e., the lowest $N$ eigenstates are occupied.
Then increase the particle number to $N'$, but choose
\ben
\gamma = N/N' \leq 1
\een
i.e., $\hbar$ is reduced in proportion to the increase in particle number.
Of course, there will now be $N'$ particles in our well, so define
\ben
\tilde\n\g(x) = \frac{N}{N'}\,\n\g(x)
\een
as a renormalized density, whose particle number matches the original value at $\gamma=1$.
This process is illustrated in Fig ~\ref{f:TFlimit}, where we plot renormalized densities
for several particle numbers $N'$, and the TF result in the same potential as used 
for Fig.~\ref{f:illus}.
One can see how our procedure reproduces the TF density, almost.  As $N'$ grows, the
oscillations in the interior of the box become smaller (with an amplitude of $O(1/N')$),
while at the edges (within O($L/N'$) of the wall), the exact density always drops to
satisfy the boundary condition, while the TF density does not.
So we also define a moving average of a function of $x$ as:
\ben
\langle \n(x) \rangle_{\Delta x} = \int_{x-\Delta x/2}^{x+\Delta x/2} dx' \n(x')/\Delta x.
\een
The length scale of the  moving average is chosen to be much larger than that of the 
quantum oscillations of the exact density and of the boundary region at the wall, 
but still vanishes as $\gamma\to 0$.
Then, finally,
\ben
\lim_{\gamma\to 0}
\langle\tilde\n_{\gamma(N',\mu)}(\mu,x) \rangle_{\sqrt{\gamma}L} = n\TF_N (x).
\een

Thus we see that the TF density in a given problem is the limit as
$\gamma\to 0$, but the convergence is highly non-uniform.  At the
walls, the true density is always zero, but the TF density is finite.
There are likely many other averaging procedures, such as taking the
limit as a finite temperature vanishes, which can be used to define the
limit, but the current one is suffificient for our present purpose.
Similarly, we define
$\tilde{t}\g(x)\equiv\gamma^3t\g(x)$.

\sec{Leading corrections to local approximations}
\label{s:leadcor}

In this section, we derive the leading corrections in $\gamma$ semiclassically,
using only elementary techniques, for the sake of transparency.  
The first such derivation was by Kohn and Sham,\cite{KSb65}
using a very elegant analysis of the properties of the Green's function
in the complex plane.  We include an appendix in which we also
derive our results via this method.  In this section, we simply
derive formulas for large $N$ and extract the $\gamma$-dependence
from such formulas.

\ssec{Density}

As in Sec.~\ref{s:clascont}, 
the density of $N$ non-interacting fermions is approximated 
by the sum of the squares of the WKB orbitals,  normalized and
satisfying the boundary conditions.  Because we are deriving
the leading corrections, we carefully normalize here:
\ben
\phi_j\WKB(x) = \sqrt{\frac{2}{k_j(x)T_j}}\sin{\theta_j(x)}\,,
\een
where the normalization constant is found by ignoring 
the oscillating term.  Define
\ben
\tau_j(x) = \int_0^x \frac{dx}{k_j(x)}\,,~~~~~~T_j=\tau_j(L)\,,
\een
the classical time for a particle in level $j$ to travel from
$0$ to $L$.
Thus, in WKB theory, the density is approximated by
\ben\label{nsum}
\n\WKB(x) = 
      \sum_{j=1}^N\frac{1-\cos{2\theta_j(x)}}{k_j(x)T_j}\ .
\een
Performing the sum exactly yields nothing other than the standard WKB 
approximation to the density as the sum of WKB densities\cite{MP56}.
However, such an approximation is inconsistent, since it retains
the discrete nature of the eigenvalues, and will not yield a
sum with a well-defined expansion in $\hbar$.
We wish to develop approximations that are smooth in $\hbar$, and
yield the exact approach to the classical continuum limit, ignoring the
discrete nature of the eigenstates, i.e., we wish to build in
the smooth envelope of functions such as the density.  At the end,
we requantize our results, and find more accurate summations
than WKB, even for $N=1$.

Begin with the smooth (non-oscillating) contribution.  We use 
the Euler-Maclaurin formula in the following form:
\ben\label{EM}
\sum_{j=1}^N f_j = \int_{\half}^{N+\half} dj f_j 
                  - \frac{1}{24}(f'\F-f'\m) 
                  + \O(f'')\,,
\een
where prime denotes a derivative with respect to $j$ and a subscript
$F$ denotes evaluation at the upper limit of the integral, $j\F=N+1/2$,
while a subscript $m$ denotes evaluation at the lower limit, $j=1/2$.
This is an expansion for sums in the same parameter as for the WKB eigenfunctions,
i.e., gradients of the potential. We retain only the first two terms, 
consistent with our WKB approximation for the orbitals.

To expand the sum of the smooth terms in such powers,
we need to relate the level index $j$ with the energy
in a continuous fashion.
Write the WKB quantization condition as
\ben\label{wkbqc}
\Theta_j
= \theta(\tilde{\e}_j,L)
= j \pi\,,~~~~~~~~~j=1,2,...
\een
which defines $\tilde{\e}_j$, the WKB eigenvalue implicitly.  
Then differentiation yields:
\ben
\Theta'_j = T_j\, \tilde{\e}_j' = \pi\,,
\een
where $\mu\sc=\tilde{\e}\F=\tilde{\e}_{N+1/2}$.
This allows us to apply Eq.~(\ref{EM}) to
the smooth contribution from Eq.~(\ref{nsum}).
Define 
\ben
\kappa_j(x) = \frac{1}{k_j(x)\, T_j}\,,
\een
which has units of inverse length and whose $j$-dependence is
typically weak, vanishing entirely for a flat box.  Then
\ben\label{TFj}
\sum_{j=1}^{N}\kappa_j(x)
\approx \int_{\half}^{N+\half}dj \kappa_j
= \int_{k\m}^{k\F}\frac{dk}{\pi}\,,
\een
where we have neglected terms that contain derivatives of $\kappa_j$
at the end points.  Thus
\ben\label{TFsmooth}
\sum_{j=1}^{N}\kappa_j(x)
 \approx \left(k\F(x)-k\m(x)\right)/\pi \,,
\een
where the quantities in the integrand of Eq.~(\ref{TFj})
depend on $j$ in a continuous manner, 
$k\F(x) = \sqrt{2\mu\sc(x)}$ and $\mu\sc$ satisfies 
the quantization condition in Eq.~(\ref{wkbqc}) 
with $j=N+1/2$, while $k\m(x)$ is the same but with $N=0$.

The oscillating term in Eq.~(\ref{nsum}) is more delicate.  
For each $x$, we expand $\theta_j(x)$
about its value at the Fermi level linearly
\ben\label{expandth}
\theta_j(x) = \theta\F(x) - (j\F-j)\,\alpha\F(x) + \dots\,,
\een
where
\ben 
\alpha_j(x)=\theta'_j(x)=\pi\,\frac{\tau_j(x)}{T_j}\ .
\een
If we truncate at this level, and
use the geometric sum defined by
\ben
h^{(0)}(z)=\sum_{j=1}^N z^j = z\, \frac{1-z^N}{1-z}\,,
\een
and using $z=\exp[i\, 2\alpha\F(x)]$, 
we find
\ben\label{sumosc}
-\sum_{j=1}^N\frac{\cos{2\theta_j(x)}}{k_j(x)T_j}=\frac{k\m(x)}{\pi}
-\frac{\kappa\F(x)\sin{2\theta\F(x)}}{2 \sin{\alpha\F(x)}}\ .
\een
The first term here exactly cancels the second term of the smooth contribution
in Eq.~(\ref{TFsmooth}). It is found from performing the geometric sum,
using Eq.~(\ref{expandth}) to undo the linear approximation at the lower
end of the sum in Eq.~(\ref{sumosc}).
Hence, the semiclassical density is
\ben
\n\sc(x) = \n\s(x) + \n\osc(x)\,, 
\een
where $s$ denotes the smooth term,
\ben
n\s(x) = \frac{k\F(x)}{\pi}\,,
\een
and $osc$ the oscillating contribution, defined to have zero 
moving average as $\gamma\to 0$,
\ben\label{nosc}
\n\osc(x)=
-\frac{\sin{2\theta\F(x)}}{2 T\F k\F(x)\sin{\alpha\F(x)}}\ .
\een

Note how completely different these corrections are from those
of Eq.~(\ref{ngea}). This is a central result of this work.

In general, the smooth term does {\em not} match
that of TF theory, because it is evaluated at $N+1/2$, not $N$.
This is not an artifact of Eq.~(\ref{EM}), but reflects the $1/2$
electron loss of density in the quantum correction.
We write Eq.~(\ref{EM}) in a form that avoids terms with
$f\m$ and $f\F$, but the $1/2$ term is independent of any particular choice.

Note also that our semiclassical density is
not normalized, in general. For a flat box, the quantization \emph{does} imply
correct normalization, but not more generally. It is straightforward to
find slightly modified definitions of $\theta(x)$, etc., that both normalize
the density and satisfy the boundary conditions (i.e., $\Theta=j\pi$), but
we choose to retain this error as a measure of the error in our 
semiclassical approximations.  We discuss this fact 
and assess the error in Sec.~\ref{ssec:en}.

Finally, we can rewrite the result as:
\ben\label{nsc}
\n\sc(x)=\frac{k\F(x)}{\pi}\left[1-\eta(x)f(\alpha\F)w(\theta\F)\right]\,,
\een
where  
$f(\alpha)=1/\sin\alpha$,
$w=\sin(2\theta)$, and
\ben
\eta(x)= \frac{\pi}{2k\F^2(x)T\F} = \frac{\hbar \omega\F}{8(\mu\sc -v(x))}
\een
is the small parameter, once $x$ is not too close to the wall, and
$\omega\F=2\pi/T\F$ is the classical frequency of collisions with the walls at
the Fermi energy.   To show 
the $\gamma$-dependence explicitly, replace
$N$ by $N/\gamma$ and write 
\bea
\tilde{\n}\sc{,\g}(x) &\equiv& \gamma\,\n\sc{_,}_\frac{N}{\gamma}(x)\\
&=& \gamma\left( \frac{k\F{_,}{\g}(x)}{\pi} - 
  \frac{\sin 2\theta\F{_,}{\g}(x)}{2T\F{_,}{\g}\,k\F{_,}{\g}(x) \sin\alpha\F{_,}{\g}(x)} \right)\,,\nonumber
\eea
where $F,\gamma$ denotes evaluation at $N/\gamma + 1/2$.

\ssec{Kinetic energy density}

A similar analysis can be applied to the kinetic energy density,
but must be done more carefully:
\ben\label{tsum}
t(x) =\sum_{j=1}^N \frac{k_j^2(x)}{2}|\phi_j(x)|^2
     \approx \sum_{j=1}^N\frac{\xi_j(x)}{2}(1-\cos{2\theta_j(x)})\,,
\een
where $\xi_j(x)=k_j^2(x)\, \kappa_j(x)$.
First we evaluate the sum over the smooth contribution 
using the same logic as for the smooth piece of the density.
Applying Eq.~(\ref{EM}) we obtain
\ben\label{ts}
\sum_{j=1}^{N}\frac{\xi_j(x)}{2}
\approx
\left[ \frac{k_j^3}{6\pi} - \frac{\xi_j'}{48}\right]^{N+\half}_{\half}\,
+ O(\xi'')\ .
\een
We know that the contributions from the lower end will be cancelled by
analogous contributions in the oscillating piece.  To evaluate that,
we define:
\ben
h^{(p)}(z)=
\sum_{j=1}^N (j\F-j)^p\, z^j = z\, h^{(p-1)}{'}(z) - j\F\, h^{(p-1)}(z)\,,
\een
where $h^{(p)}{'}(z)=dh^{(p)}/dz$.

Each term has many terms, but only those
containing a $z^N$ will contribute to our answer, because 
when we insert Eq.~(\ref{expandth}) into Eq.~(\ref{tsum}), the
prefactor contains $z^{-N}$.
Then
\ben
t\osc(x) = -\half \Re\ \left\{e^{2i(\theta\F-j\F\alpha\F)}
\left[ \xi\F h^{(0)} -\xi'\F h^{(1)} + \half\xi''\F h^{(2)}\right]_{z=e^{2i \alpha\F}} \right\}.
\een
Evaluating term by term yields
\ben\label{tosc}
t\osc(x)= \frac{1}{16}\left\{\xi\F\, f \frac{\partial^2 w}{\partial\theta^2}
+\xi\F' \frac{\partial f}{\partial \alpha}\frac{\partial w}{\partial\theta}
+\half \xi\F''\, \frac{\partial^2 f}{\partial \alpha^2} w\right\}\ .
\een
The derivatives of $\xi$ w.r.t. $j$ can be written as
\ben\label{xip}
\xi_j'= 2\pi \xi_j^2/k_j^3 +k_j^2 \kappa_j'
\een
and
\ben\label{xipp}
\xi_j''= 2\pi^2 \xi_j^3/k_j^6 + 6\pi\kappa'_j/T_j
+k_j^2\kappa_j''\ .
\een
We now drop all derivatives of $\kappa_j$, because they vanish in the
flat box limit, yielding:
\ben\label{tsc}
\tilde{t}\sc(x)=\frac{k\F^3}{6\pi}\left[1 + \frac{3}{4}\eta f\frac{\partial^2 w}{\partial\theta^2} 
+\eta^2\left(3\frac{\partial f}{\partial\alpha} \frac{\partial w}{\partial\theta}-1\right)
+ 3\eta^3\frac{\partial^2f}{\partial\alpha^2} w\right]\ . 
\een
Again, the explicit $\gamma$-dependence of this formula is found by 
replacing $N$ by $N+1/2$.  As we shall show in later sections, 
this result is less well-behaved than that for the density. 
For example, when the potential is non-uniform, the semiclassical
kinetic energy density of Eq.~(\ref{tsc}) incorrectly 
fails to vanish at the edges.  

To overcome this failure, we define the edge as being those values of $x$ 
up to some fraction $\beta$ of a period of the classical phase:
\ben
\theta\F(x_\beta) = \beta\,\pi\,,
\een
such that the edge region is $x<x_\beta$ and $x>L-x_\beta$.
We choose $\beta=1/4$ (and the interior is all the rest).  
This mimics the approach used in boundary-layer theory for differential equations,
which can be applied to the $\hbar$-expansion of the individual
levels.  One constructs approximations that are correct to a given
order in the asymptotic expansion in each region separately, and hopes
to find a middle region where they match, yielding a 
solution with uniform convergence properties,\cite{BO99} i.e., with the correct
asymptotic expansion for any $x$.  The only difference here is that
we are applying these ideas to the sum of levels, not the individual levels
themselves.

Our final semiclassical approximation for the KED is
\ben\label{tsco}
t\sc(x) = 
\begin{cases} 
\tilde{t}\sc(x)      & \mbox{if}~~~~~~~ x_\beta < x < L-x_\beta\,, \\
\tilde{t}\unif\sc(x) & \mbox{else}\,,
\end{cases}
\een
where
\bea
\tilde{t}\unif\sc(x) &=& \frac{(k\F\unif)^3}{6\,\pi}
                      -  \frac{(k\F\unif)^2\sin(2\,k\F\unif x)}{4\,L\sin(\pi x/L)}\nonumber\\
                     &-& \frac{\pi\, k\F\unif \cos(\pi x/L) \cos(2\,k\F\unif x)}{4\,L^2\sin^2(\pi x/L)}\nonumber\\
                     &-& \frac{\pi^2\sin(2\,k\F\unif x)}{8\,L^3\sin(\pi x/L)}
                         \left(\half - \frac{1}{\sin^2(\pi x/L)} \right) .
\eea
Hence,  
inside the edge region we approximate the KED by  
$\tilde{t}\sc\unif(x)$ meaning that we evaluate Eq.~(\ref{tsc}) with the 
local Fermi wavevector for a uniform potential, i.e., replacing $k\F$
everywhere by $k\F\unif=\sqrt{2\mu\sc}$ and defining
all other quantities based on that. In particular the classical phase 
and transit time become linear in $x$, as $k\F\unif$ for the uniform system is 
independent of $x$.  Hence, the boundary conditions will always be satisfied, 
no matter what $v(x)$ is.
Outside that region, i.e., in the interior of the box ($x_\beta < x < L-x_\beta$), 
the nonlocal $k\F(x)$ is used. 
\begin{figure}[htb]
\begin{center}
\includegraphics[angle=0,height=6cm]{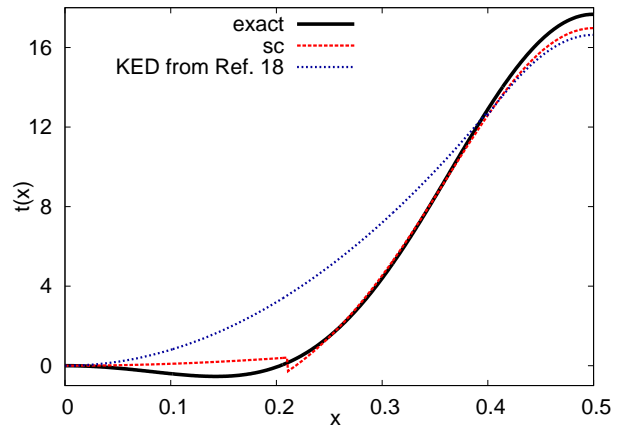}
\end{center}
\caption{Exact and approximate ground-state KEDs
for $v(x)=-12\sin^2(\pi x)$, where $0\le x\le 1$.
The lowest eigenvalue is $\e_0=-4.27$ and $\mu\sc=5.52$.}
\label{f:teva_D12N1}
\end{figure}
We illustrate our approximations and the exact KED
for a single-well potential $v(x)=-12\sin^2(\pi x)$ 
within box boundaries in Fig.~\ref{f:teva_D12N1}.
Note that our present approximation of Eq.~(\ref{tsco}) is substantially
more accurate for both the edge region and the interior than 
the previously derived KED of Ref.~\onlinecite{ELCB08}.
In the next section we discuss this fact more quantitatively.

\sec{Properties}
\label{sec:props}

We next test our approximations, to demonstrate both their accuracy and
that they have the properties claimed for them.  We begin with
several integrated quantities, mostly energies.

\ssec{Energies and normalization}
\label{ssec:en}

The LPA yields densities that are local in the
potential, with the exception of the value of the chemical potential, which
must be determined globally.  The quantum corrections depend on several other
terms, such as $\theta\F(x)$ and $T\F$, which are still simple functionals
of the potential, but distinctly non-local, depending on integrals over $v(x)$.
To test the integrated properties of the density, we calculate
moments over that density.  The obvious one is the third moment, as that
is simply related to the local {\em density} approximation to $T$, 
evaluated on that density.

\begin{table}[htb]
\caption{Exact and approximate quantities for
one particle in a single-well potential
$v(x)=-10\sin^2{(\pi x)}$, $0\le x\le 1$.
$T$ is the exact kinetic energy and $n$ the exact density.}
\label{t:basicnumbers}
\begin{ruledtabular}
\begin{tabular}{ l | c c c c c }
Energy levels & $\e_1$ & $\e_2$ & $\mu$ & $\mu\sc$ & \\
\hline
& -2.71 & 14.6 & 0.637 & 6.38 & \\
\hline
\multicolumn{1}{l|}{Kinetic energy} & 
\multicolumn{2}{c}{$T$} & 
\multicolumn{3}{|c}{$T\loc[n]$} \\
\hline
\multicolumn{1}{c|}{} &
\multicolumn{1}{c}{exact} &
\multicolumn{1}{c}{sc}  & 
\multicolumn{1}{|c}{TF}   & 
\multicolumn{1}{c}{exact}   & 
\multicolumn{1}{c}{sc}\\ 
\multicolumn{1}{c|}{} &
\multicolumn{1}{c}{5.07} &
\multicolumn{1}{c}{5.02}  & 
\multicolumn{1}{|c}{2.31}   & 
\multicolumn{1}{c}{4.93}   & 
\multicolumn{1}{c}{5.07} 
\end{tabular}
\end{ruledtabular}
\end{table}
We choose a standard potential, $v(x)=-10 \sin^2 \pi x$ in a box of length 1,
and insert one particle.   Both exact and approximate results are given
in Table \ref{t:basicnumbers}.  

First note that the TF result, $T\TF$, is about 50\% too small, compared to the exact answer, $T$.  This is the result of minimizing the energy using LDA
as in the first term of Eq.~(\ref{Tgea}).

We measure the quality of the TF density and our semiclassical density 
by evaluating the LDA kinetic energy on those densities, i.e., 
$T\loc[\n\TF]=T\TF$ and $T\loc[\n\sc]$, where the point of reference is the
LDA kinetic energy evaluated on the exact density, $T\loc[\n]$.
But the TF result remains about 50\% too small compared to $T\loc[n]$.  
However, the LDA on our semiclassical density, $T\loc[\n\sc]$, 
yields an energy only 3\% too large, i.e., reducing the error by 
about a factor of 20.  

To test our semiclassical kinetic energy, $T\sc$, we compare with
the exact value, $T$, and find an error of only 0.9\%
too small, i.e., 50 times better than $T\TF$.
Thus the semiclassical results are more than an order of magnitude
better than bare DFT results, because they include quantum oscillations.
In fact, the LDA kinetic energy evaluated on the exact density,
$T\loc[n]$, yields only a 2.7\% underestimate,
showing that local approximations do
much better on accurate densities, but still not as well as our
direct approximation, $T\sc$.

These systems do not appear to be particularly {\em semiclassical}:
the potential is not flat nor is the particle number or index high.
We can analyze the source of this accuracy by expanding
integrated quantities in powers of $\gamma$ about 0:
\ben\label{Tgammaexp}
T(\gamma) = T^{(0)} + \gamma\, T^{(1)} + \gamma^2\, T^{(2)} + \ldots
\een
For the kinetic energy, from the previous discussion:
\ben
T^{(0)} = T\TF 
\een
while our derivation should yield:
\ben
T^{(1)}=T\sc^{(1)}\ .
\een
These results should hold for {\em both} the local approximation applied
to the exact density (and so test our semiclassical density) and the
exact kinetic energy (and so test our semiclassical kinetic energy density).
\begin{figure}[htb]
\begin{center}
\includegraphics[angle=0,height=6cm]{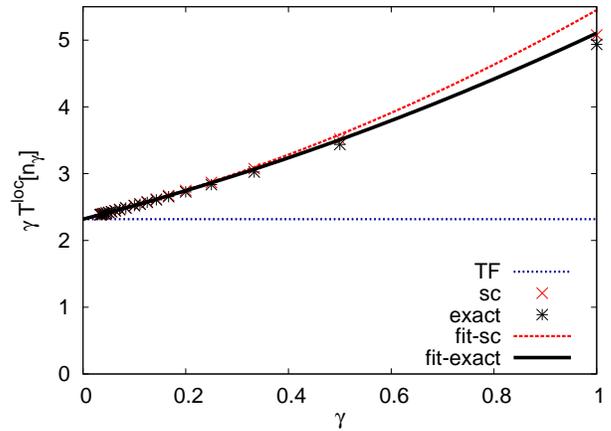}
\end{center}
\caption{LDA kinetic energy multiplied by the scale factor $\gamma$
for different $\gamma$ evaluated on $\n\g\TF(x)$, 
our semiclassical density $\n\sc{,\g}(x)$, 
and the exact density for a single-well potential 
$v(x)=-10\sin^2{(\pi x)}$.}
\label{f:Tloc}
\end{figure}
In Fig.~\ref{f:Tloc}, we study the $\gamma$-dependence of $T\loc[\n]$ 
applied to various densities for a generic well.  Clearly TF gives the
$\gamma=0$ value, while the semiclassical density includes the correct
linear contribution, and is quite accurate for higher-order contributions.
We also note that inclusion of the linear term greatly improves over the
TF result, but that the LDA kinetic energy evaluated on our semiclassical 
density is even more accurate still.

Because our expansion is in powers of $\hbar$, we expect that it 
is asymptotic, just as the WKB expansion is.\cite{BO99}  Thus, for fixed $N$
and $\gamma$,  inclusion of additional coefficients in the expansion
will eventually {\em worsen} the result.  We can see this in Table~\ref{t:Tgammaexp},
where the error of our semiclassical result, $|T\sc-T|$, at $\gamma=1$ is smaller
than the error in the quadratic coefficients, $|T\sc^{(2)}-T^{(2)}|$, 
and thus cannot be explained in terms of its approximation 
to that (or any higher) coefficient.
\begin{table}[htb]
\caption{Coefficients of $\gamma$-expansion in Eq.~(\ref{Tgammaexp}) of the exact
and semiclassical kinetic energy, $T^{(i)}$ and $T\sc^{(i)}$, and the
values $T$ and $T\sc$ at $\gamma=1$ for
$v(x)=-10\sin^2{(\pi x)}$, where $0\le x\le 1$.}
\label{t:Tgammaexp}
\begin{ruledtabular}
\begin{tabular}{ l | c c c c c c }
$N$ & $T^{(0)}$ & $T^{(1)}$ & $T^{(2)}$ & $T\sc^{(2)}$ & $T$ & $T\sc$\\
1 &   2.31 &  2.05 &  0.614 &  0.900 &   5.07 &   5.02\\
2 &  13.4  &  9.78 &  1.69  &  1.62  &  24.9  &  24.7\\
\end{tabular}
\end{ruledtabular}
\end{table}
On the other hand, the asymptotic expansion with just the first 
few terms in Eq.~(\ref{Tgammaexp}) becomes accurate very rapidly as $N$ increases.
Compare the relative error of the quadratic coefficients of about $50\%$ for $N=1$ 
with roughly $4\%$ for $N=2$.  The error dropped by about an order
of magnitude as the number of particles increases by one.
To understand why that is so, we next consider the $N$-dependence of
each contribution, where $N$ is now the number of particles at $\gamma=1$.
As $N\to\infty$, the box must appear flat.  Evaluating the local
approximation on the flat box density yields:
\ben\label{Tlocgamma}
T\g\loc[\n] = \frac{\pi^2 N^3}{6 L^2} 
\left[1 + \frac{9}{8}\left(\frac{\gamma}{N}\right) 
        + \frac{3}{8}\left(\frac{\gamma}{N}\right)^2\right]\,,~~(flat)
\een
fixing the first three coeffcients with the values above, and the rest to vanish.
The corrections to this flat limit can only involve powers of $1/N$, which we
can either derive or find numerically.

Since the leading term is given by TF theory, if we expand the potential in the
box in a power series around its average value, we find
\ben
T\TF = \frac{\pi^2 N^3}{6 L^2} \left(1+ \frac{3\,{\overline{\delta v^2}}}{(\pi {\overline n})^4} +\ldots\right)
\een
where 
$\overline n = N/L$ and
\ben
{\overline{\delta v^2}}= \int_0^L dx\, (v(x)-\overline v)^2/L 
\een
with $\overline v$ the average of the potential over the well.  
For our shape, ${\overline{\delta v^2}}=D/8$, yielding a TF value of $2.31$, as in the figure.
More importantly, we see that the leading correction to the flat box result is
$O(1/N^4)$.  Similarly, we find by fitting, that
\ben
T^{(1)}=\frac{3\pi^2 N^2}{16 L^2} \left(1+ \frac{a}{N^2} + \frac{b}{N^4} + \ldots\right)
\een
and is given exactly by the semiclassical approximation.  For the specific
choice of potential $v(x)=-10\sin^2(\pi x)$ the coefficients are
$a=0.38$ and $b=-0.26$.
Finally,
\ben
T^{(2)}=\frac{\pi^2 N^2}{16 L^2} \left(1+ \frac{c}{N^4} + \ldots\right)\,,
\een
but the coefficient $c$ is {\em not} given correctly by the semiclassical
approximation. For $v(x)=-10\sin^2(\pi x)$ the exact value is $c=0.42$,
whereas the semiclassical approximation gives about half that value.
Thus, all corrections to the flat results vanish rapidly as $N$ increases, and
the errors of the first few semiclassical terms in the expansion become much smaller,
leading to a much more accurate value at $\gamma=1$.

\begin{table}[htb]
\caption{Local approximation to the kinetic
energy evaluated on the TF, semiclassical, and exact density, 
and the kinetic energy from direct integration
of the semiclassical and exact KED, all relative to the 
flat box value for $N$ particles in
a single-well potential $v(x)=-10\sin^2{\pi x}$, $0\le x\le 1$.
The errors of our semiclassical result with respect to the exact
result are denoted by $\Delta \rm{sc}$.}
\label{t:Ndep}
\begin{ruledtabular}
\begin{tabular}{@{\extracolsep{\fill}}c||r c c | c || c c |c}
\multicolumn{5}{c||}{$\Delta T\loc[n]$} &
\multicolumn{3}{c}{$\Delta T$}\\
\hline
$N$ & TF & sc & exact & $\Delta$sc & sc & exact & $\Delta$sc \\
\hline
1 &   -1.8 &  0.96 &  0.82 &  0.14 &  0.09 &  0.13 & -0.04\\
2 &   -8.3 &  0.89 &  0.92 & -0.03 &  0.08 &  0.24 & -0.16\\
4 &  -32   &  0.77 &  0.78 & -0.01 &  0.09 &  0.14 & -0.05\\
6 &  -70   &  0.72 &  0.73 & -0.01 &  0.07 &  0.09 & -0.02\\
8 & -123   &  0.70 &  0.70 &  0.00 &  0.06 &  0.07 & -0.01\\
\end{tabular}
\end{ruledtabular}
\end{table}

In Table \ref{t:Ndep}, we list the various kinetic energies as functions of $N$ 
for our well.  Because the errors vanish so rapidly, we subtract the energies
of the uniform system, as in:
\ben
\Delta T = T - T\unif
\een
and likewise for $\Delta T\loc[n]$.
These differences could also be thought of as the change in energy due to turning on the
well in the bottom of the box, analogous to the change in energy when atoms
form a molecule.  We see that our approximations become very accurate very quickly,
and converge as $1/N^2$.

The quantum correction yields a density that is {\em not} normalized.  This is because
the requirement in Eq.~(\ref{wkbqc}) that the phase vanishes at both 
$x=0$ and at $x=L$ is used to determine $\mu\sc$, 
not simple normalization.  Of course, the
error vanishes rapidly as $N\to\infty$; for $N=1$, it is 
$\Delta N=4\times 10^{-2}$ and for $N=2$, $\Delta N=6\times 10^{-4}$.
One can easily imagine schemes
that patch this failure up, but we prefer to leave it as a measure of
the overall error in the approximation.

Since our formulas reduce to the exact results for a uniform potential, 
more generally, they should preserve these good features for 
a slowly-varying potential.  We have applied our density formula 
to many examples, and {\em almost always}
found it to be remarkably accurate.   This is because of its
excellent formal properties, and because we capture the leading correction
to the LPA in a well-defined (albeit asymptotic) series.  Most importantly,
it appears that the conditions of application, $\mu\sc$ above $v$ everywhere,
imply that these leading corrections {\em always} improve over the
dominant contribution.

\ssec{Uniform convergence}
\label{ssec:unifcon}
While the most important aspect of our work is the recovery
of the leading asymptotic corrections to TF for the energies,
the detailed spatial dependence is also important for understanding
how this is achieved, and also for understanding the strengths and
weaknesses of this approach.

Our semiclassical approximations 
are exact in the case of a uniform potential, 
where they yield the simple formulas:
\ben
n\unif\g(x)=\N\left(1-\frac{\sin{(2\pi\N x)}}{2\N\sin \pi x} \right)
\een
and
Eq.~(\ref{tsc}) with 
$f = 1/\sin(\pi x)$,
$w = \sin(2\pi\N x)$, and 
$\N= (N/\gamma+1/2)$.
These offer some insight into the nature of the expansion.  

Consider
first the density.  For any finite value of $x$, the oscillating
contribution shrinks and oscillates more rapidly as $\N\to\infty$.
Thus, we can expand the smooth part, the prefactor of
the oscillating contribution, and the phase of the oscillation,
in powers of $1/\N$, which is linear in $\gamma$ for small $\gamma$.
On the other hand, for $\N x$ fixed, one can again expand the
density for large $\N$:
\ben
\n\g(y)= \N\left[\left(1-\frac{\sin 2\pi y}{2\pi y}\right) 
       - \frac{\pi y \sin(2\pi y)}{12\N^2} + ...\right]\,,
\een
where $y=\N x$.
The first term is precisely the profile of a semi-infinite box
at the surface.  This series is very ill-behaved for large $y$, except for the lowest-order
term.  Similar comments apply to the kinetic energy density only more so,
as several contributions {\em diverge} for small $\N x$.
Thus there are two distinct regions and limits within
the well, the interior and the edges.  

In what follows we illustrate that the error of our
semiclassical approximations converges uniformly
as $\gamma\to 0$.   We define:
\ben
\Delta n\sc(x)=\n\sc(x)-\n(x)
\een
as the error in the semiclassical density, and likewise for the
kinetic energy density.
\begin{figure}[htb]
\begin{center}
\includegraphics[angle=0,height=6cm]{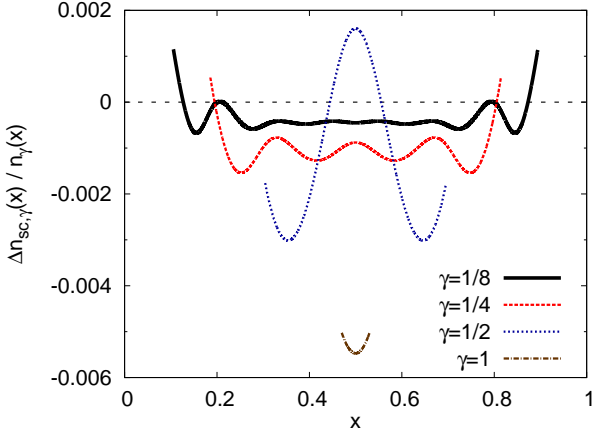}
\end{center}
\caption{Fractional error in density 
for $v(x)=-5\sin^2{(\pi x)}$; only shown in interior
($\theta\F/\pi \ge 0.7$ in left half).}
\label{f:dn_int_D5m1}
\end{figure}
\begin{figure}[htb]
\begin{center}
\includegraphics[angle=0,height=6cm]{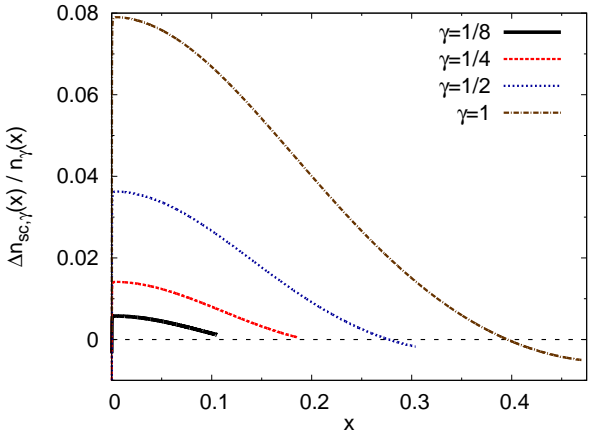}
\end{center}
\caption{Fractional error in density 
close to the edge for $v(x)=-5\sin^2{(\pi x)}$,
$\theta\F/\pi \le 0.7$.}
\label{f:dn_ext_D5m1}
\end{figure}
\begin{figure}[htb]
\begin{center}
\includegraphics[angle=0,height=6cm]{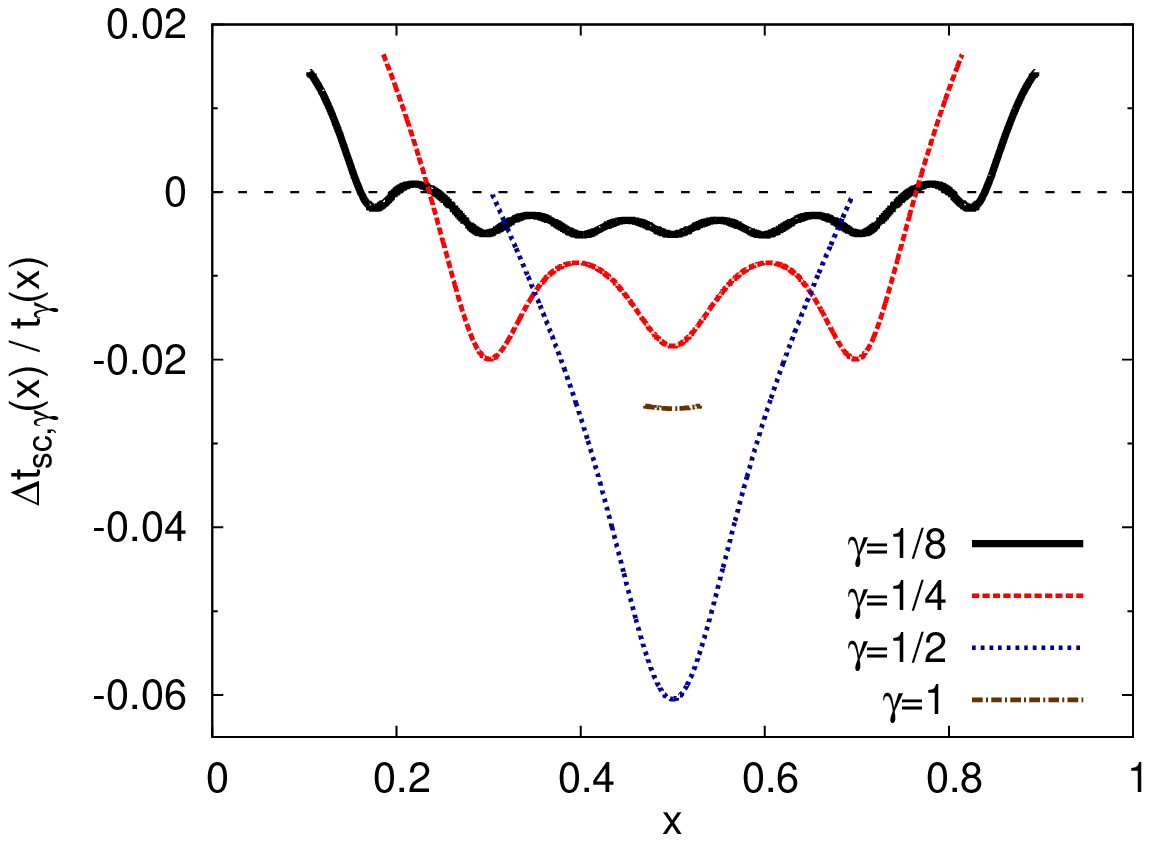}
\end{center}
\caption{Same as Fig.~\ref{f:dn_int_D5m1}, for kinetic energy density.}
\label{f:dt_int_D5m1}
\end{figure}
\begin{figure}[htb]
\begin{center}
\includegraphics[angle=0,height=6cm]{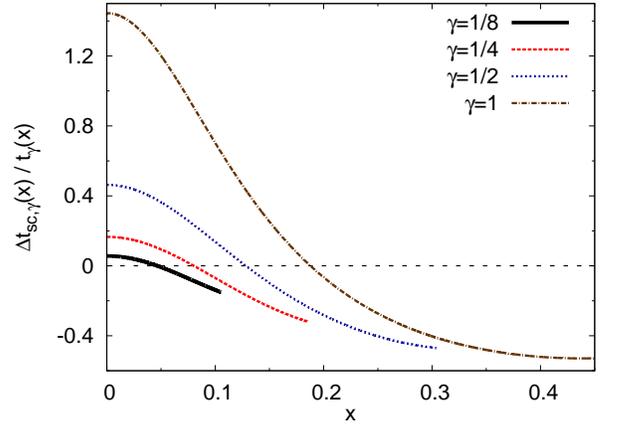}
\end{center}
\caption{Same as Fig.~\ref{f:dn_ext_D5m1}, for kinetic energy density.}
\label{f:dt_ext_D5m1}
\end{figure}
We pick a generic single-well potential 
$v(x)=-5\sin^2(\pi x)$ sufficiently close to flat
so that we are in a regime dominated by the asymptotic behavior.
For illustrative purposes we increase the extend of the edge-region 
by choosing $\beta=0.7$.
The fractional error of the density 
in the interior is shown in Fig.~\ref{f:dn_int_D5m1}.
The error converges uniformly 
throughout the interior as $\gamma\to 0$, being $\O(\gamma)$.
As shown in Fig.~\ref{f:dn_ext_D5m1}, the fractional error 
close to the edge of the box also converges uniformly, being $\O(\gamma)$
but noticeably larger.
The convergence for the KED is shown in Figs.~\ref{f:dt_int_D5m1}
and \ref{f:dt_ext_D5m1}, and has the same features as the
density, but is much larger.

\ssec{Phase oscillations}
We also check that the quantum oscillations of our
semiclassical formulas can be extracted from the 
exact results as $\gamma\to 0$.
For a fixed point $x$ we look at the difference
between the exact result and the smooth term, multiplied
by the prefactor appearing in our formula for the quantum
oscillations:
\ben\label{nqc}
d\g(x)=2T\F k\F(x)\sin{\alpha(x)}
\Delta n\s{,\g}(x).
\een
If our results are correct, as $\gamma\to 0$, this becomes 
a simple function of $S\F(x)/\gamma$, for {\em any} values of $x$ and
$\gamma$, specifically $-\sin (2S\F(x)/\gamma)$.  
\begin{figure}[htb]
\begin{center}
\includegraphics[angle=0,height=6cm]{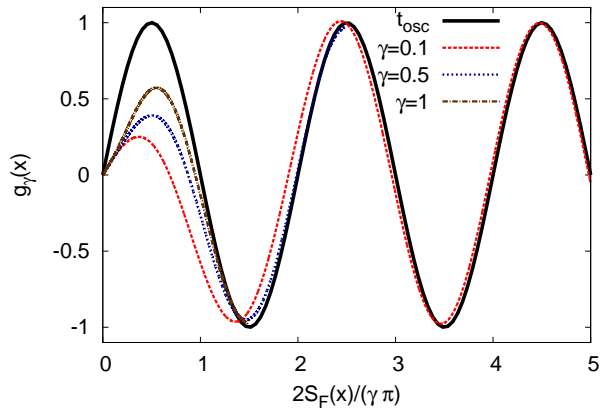}
\end{center}
\caption{Leading correction to the kinetic energy density
for $\gamma=1, 0.5, 0.1$ and $v(x)= -10\sin^2(\pi x)$.} 
\label{f:tqc1}
\end{figure}
The same analysis applies to the kinetic energy density, where 
we define:
\ben\label{tqc1}
g\g(x)=\frac{4T\F\sin\alpha(x)}{k\F(x)}\Delta t\g\TF\,,
\een
In Fig.~\ref{f:tqc1}, we plot $g\g(x)$ for several values of $\gamma$,
as a function of $2\theta\F(x)/\pi$, finding results converging to
$-\sin 2\theta\F(x)$, as predicted by the leading term of Eq.~(\ref{tosc}).

\ssec{Evanescent regions}

\begin{figure}[htb]
\begin{center}
\includegraphics[angle=0,height=6cm]{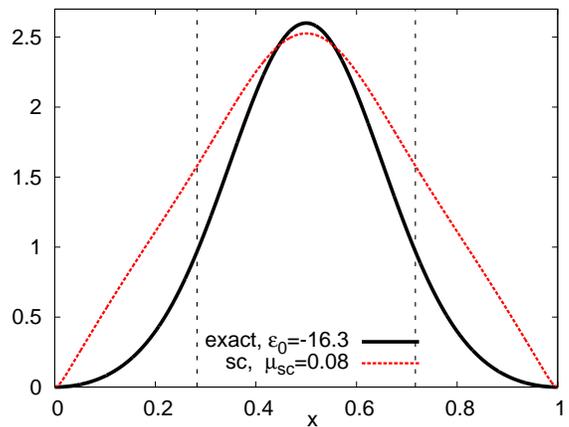}
\end{center}
\caption{Exact and approximate ground-state densities 
for $v(x)=-27\sin^2(\pi x)$, where $0\le x\le 1$.
The lowest eigenvalue is $\e_0=-16.3$ and $\mu\sc=0.08$.
The position of the turning points is indicated by dashed lines.}
\label{f:neva_D27N1}
\end{figure}

\begin{figure}[htb]
\begin{center}
\includegraphics[angle=0,height=6cm]{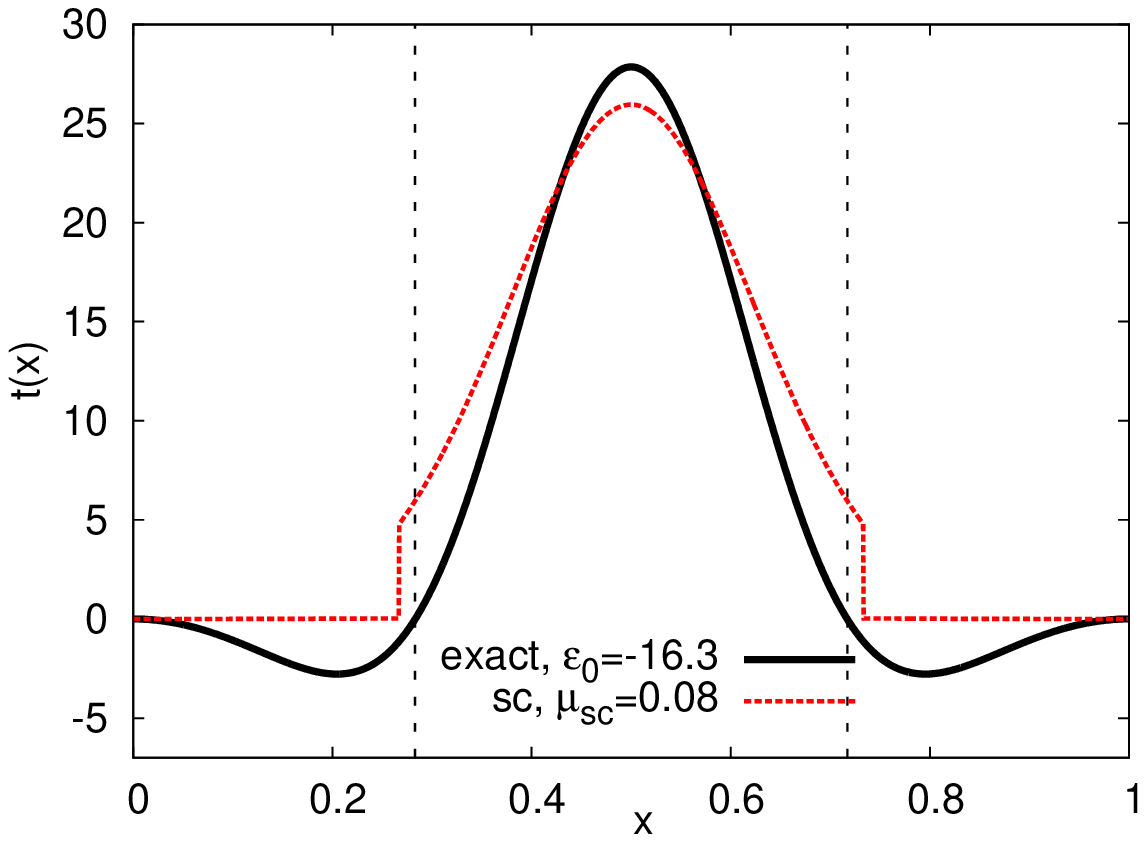}
\end{center}
\caption{Exact and approximate ground-state KEDs
for $v(x)=-27\sin^2(\pi x)$, where $0\le x\le 1$.
The lowest eigenvalue is $\e_0=-16.3$ and $\mu\sc=0.08$.
The position of the turning points is indicated by dashed lines.}
\label{f:teva_D27N1}
\end{figure}

The only condition on the applicability of
our approximations is that $\mu\sc > v(x)$ for all $x$, 
But $\mu\sc$ is between the highest occupied and lowest unoccupied level,
and so for many well-depths, it can be the case that the HOMO 
has turning points while the condition is still valid.
The starkest example is for $N=1$, since beyond those turning points,
the only occupied orbital is evanescent.  Yet our approximations
can still be applied, even though they contain only trigonometric
functions of the phase, and no decaying exponentials, and still
yield {\em highly accurate} results.  
In Figs.~\ref{f:illus}
and \ref{f:teva_D12N1} we show results for 
a well depth of $12$, for which the lowest eigenvalue is 
$\e_0=-4.27$ and $\mu\sc=5.52$, and the turning points
are located at around $x=0.2$ and $x=0.8$.
Eventually the quadratic approach of the semiclassical density 
near the wall of the box mimics the exponential decay of the true density.
Even the KED truncated by our method, only misses the negative contribution,
which largely cancels the error in the interior.
The results for this well remain remarkably accurate.
As the semiclassical chemical potential $\mu\sc$ approaches $v\max$ 
the validity of our approximation breaks down.  We simulate such a 
situation in Figs.~\ref{f:neva_D27N1} and \ref{f:teva_D27N1}
by choosing a well depth of $27$, such that $\mu\sc$ is only
slightly above $v\max$. 
This is the worst qualitative breakdown of our approximation, yielding
the errors shown in Tab.~\ref{t:evanumbers}.
But even here, errors are $\le 20\%$.
\begin{table}[htb]
\caption{Exact and approximate quantities for
one particle in a single well
$v(x)=-D\sin^2{(\pi x)}$, where $0\le x\le 1$.}
\label{t:evanumbers}
\begin{ruledtabular}
\begin{tabular}{ l | c c c c c }
Energy levels & $\e_1$ & $\e_2$ & $\mu$ & $\mu\sc$ & \\
\hline
$D=12$ & -4.27 & 13.6 &  0.04 & 5.52 & \\
$D=27$ & -16.3 & 5.47 & -6.75   & 0.08 & \\
\hline
\multicolumn{1}{l|}{Kinetic energy} & 
\multicolumn{2}{c}{$T$} & 
\multicolumn{3}{|c}{$T\loc[n]$} \\
\hline
\multicolumn{1}{c|}{} &
\multicolumn{1}{c}{exact} &
\multicolumn{1}{c}{sc}  & 
\multicolumn{1}{|c}{TF}   & 
\multicolumn{1}{c}{exact}   & 
\multicolumn{1}{c}{sc}\\ 
\multicolumn{1}{l|}{$D=12$} &
\multicolumn{1}{c}{5.13} &
\multicolumn{1}{c}{5.18}  & 
\multicolumn{1}{|c}{2.66}   & 
\multicolumn{1}{c}{5.12}   & 
\multicolumn{1}{c}{5.33}\\
\multicolumn{1}{l|}{$D=27$} &
\multicolumn{1}{c}{5.74} &
\multicolumn{1}{c}{7.63}  & 
\multicolumn{1}{|c}{4.80}   & 
\multicolumn{1}{c}{6.42}   & 
\multicolumn{1}{c}{8.47} 
\end{tabular}
\end{ruledtabular}
\end{table}

\sec{Consequences for density functional theory}
\label{s:conseq}
This work has been confined to one-dimensional non-interacting particles
confined by hard walls.
In this section, we discuss in detail the ramifications for density functional
theory in the real world of atoms, molecules, and solids.
We divide the discussion in two:  Thomas-Fermi theory and  
Kohn-Sham theory.

We begin with Thomas-Fermi theory and its extensions.  This 
was the original density functional theory (DFT) and continues
to be used in many fields of physics.  TF theory became obsolescent
for electronic structure calculations with Kohn-Sham work, but
there has been a recent resurgence of interest in orbital-free DFT, 
with the hope of treating systems of much greater size than is
presently possible with Kohn-Sham calculations.
To do this, all that is needed is an accurate approximation to the
non-interacting kinetic energy as a functional of the density.
The original approximation using uniform gas inputs, is simply the
3D analog of our 1D local approximation used here.  Thus if our
methods could be generalized to apply to the general 3D case,
it would produce an orbital-free theory.

Perhaps the most important result of this study is to highlight 
an alternative path.  Instead of trying to find {\em density}
functional approximations, we have derived the leading corrections
in terms of the {\em potential}, a perfectly valid alternative
variable to the density.\cite{YAW04}  If general formulas (or algorithms)
could be found for finding accurate approximations
to $\n[v\s](\br)$ and $t\s[v\s](\br)$, where the subscript $s$ denotes
non-interacting, one could use them to avoid solving the Kohn-Sham
equations and evaluating any orbitals.  At each step in the iteration,
one finds $v\s(\br)$, the Kohn-Sham potential, using some standard
XC functional, and uses this to generate a new density.
When self-consistency is reached, the kinetic energy is evaluated
and the many-body energy is found in the usual way. 
A recent study\cite{GP09} shows that this procedure is correct
once both approximations are derived from the same approximate Green's
function, as ours have been.

Even before such generalizations have been found, we have been able
to use results here to deduce information on the 3D kinetic energy functional.
In 1d, our results show that the leading corrections to the
asymptotic expansion  of the kinetic energy  in powers of $1/N$ are
{\em not} determined by the gradient expansion for any finite system,
but instead are given by the quantum corrections producing quantum oscillations.
Ref.~\onlinecite{LCPB09} is a careful study of the asymptotic expansion for
the 3D kinetic energy, and showed how generalizing the gradient expansion to
ensure recovery of the asymptotic expansion greatly improved total energies
over the gradient expansion, but worsened other energies, such as those of
jellium surfaces.  This reflects the difficulty in attempting to capture
different physical limits with simple density-functional approximations.
Even our simple results cannot be easily encoded in a density functional
approximation, but are both simple and (relatively) physically
transparent as {\em potential} functionals.

Almost all modern electronic structure calculations are performed within the
Kohn-Sham formalism, which provides a set of self-consistent non-interacting
single-particle equations which reproduce the exact single-particle density
of the interacting system.  In these, the non-interacting kinetic energy is treated
exactly, and only a small contribution to the total energy, the XC
energy, is approximated as a functional of the density.  This contribution is determined
by the Coulomb repulsion, and so is a many-body effect.

So, do we learn anything from studying our little toy problems?  The answer is definitely
yes.  Our toy is perhaps the simplest possible system in which one can meaningfully
approximate a Schr\"odinger equation with its density functional analog, and make a
local density approximation.   So we learn in what limits this becomes relatively exact,
and how to find the leading corrections.  We learn the nature of these corrections (asymptotic)
and how there are mutliple length scales in the system.
While the details of these lessons depend on the functional we are approximating,
some general features of functionals and their approximations can be guessed at, and
highly useful analogies can be made.

For example, there are many ways to understand why Kohn-Sham calculations are far more accurate
than Thomas-Fermi type calculations, and our analysis produces one more.
A KS calculation, by virtue of its orbitals, produces an incredibly accurate
density, and we have seen how local-type approximations are much more powerful
on accurate densities than on self-consistent densities.  Thus not only
does a Kohn-Sham calculation approximate only a small fraction of the
total energy (the XC piece), but even that part is
much more accurately given by a local approximation by virture of the accurate
density.

The insight based on the semiclassical analysis of functionals has already led
to significant development in the Kohn-Sham XC functional.
Schwinger demonstrated\cite{Sa80,S81} that LDA exchange becomes relatively exact
for large $Z$ neutrals.
Analysis of the large-$Z$ behavior of modern exchange GGAs\cite{PCSB06,EB09}
shows that the most popular functionals all recover (to within about 20\%) the
leading corrections to the LDA asymptotic behavior of exchange for atoms.
 On the other hand, the gradient
expansion approximation, based on the slowly-varying limit, does not, being too
small by almost exactly a factor of 2.  This is entirely analogous to our problem,
in which the local approximation recovers the exact dominant term, and a decent
approximation to the next correction, but the gradient expansion worsens that
agreement.  Since such functionals are tested on exchange energy of atoms, and
these cannot be accurate without accurate asymptotic values, this is vital
for recovering accurate thermochemistry, which requires accurate atomic energies.
On the other hand, bond lengths depend only on small variations in the energy
when the bond distances is varied slightly around its equilibrium value, and
so do not require accurate energies of isolated atoms. These can be improved
upon over regular GGAs by restoring the true gradient expansion and
ignoring the asymptotic limit.  The recent
PBEsol functional\cite{PRCV08} does exactly this for exchange
and produces better lattice parameters for many solids.

\sec{Summary}

We have presented a fuller and more precise account of the results originally
shown in Ref.~\onlinecite{ELCB08}.
Kohn and Sham\cite{KSb65} produced asymptotic expansions for the interior, exterior,
and turning point regions of the density.  In Ref.~\onlinecite{ELCB08}, we presented a uniform
approximation for the interior region, but only an asymptotic approximation
for the kinetic energy density.  Here, by analyzing the breakdown of the
method for the boundary regions, we have produced a (nearly) uniform approximation
to the kinetic energy.

This work was supported by NSF under grant number CHE-0809859. 
We also acknowledge support from the KITP under 
grant number PHY05-51164.

\appendix
\section{Derivation of the semiclassical density and kinetic energy density 
in the complex plane}
The semiclassical corrections were derived from a contour integral over
the semiclassical Green's function in Ref.~\onlinecite{ELCB08}, 
and we give a fuller account here.
The method is well-described in Ref.~\onlinecite{KSb65} 
but we go beyond the aims there, since we
require our solution to be uniformly asymptotic, not just producing 
the correct asymptotic expansion in the interior and we extract also
the kinetic energy density.  We are also 
treating box boundary conditions, rather
than the turning points discussed there.
Begin with the diagonal Green's function 
\ben
\label{e:gf-def}
g(x,\e)=\frac{2\, \phi_{L}(x)\, \phi_{R}(x)}{W(\e)}\,,
\een
where 
$W(\e)=\phi_{L}(x)\partial\phi_{R}(x)/\partial x
      -\phi_{R}(x)\partial\phi_{L}(x)/\partial x$ 
is the Wronskian,
and approximate the two independent solutions $\phi_{L}(x)$ and $\phi_{R}(x)$
via the WKB wavefunctions satisfying the boundary conditions:
\bea
\label{wkb}
\phi\WKB_{L}(x) & = & \sin[\theta(x)]/\sqrt{k(x)}\,,\\
\phi\WKB_{R}(x) & = & \sin[\theta(L-x)]/\sqrt{k(x)}\,, 
\eea
yielding
\ben
\label{gf}
g\sc(x,\e) = \frac{\cos{\Theta}-\cos{\left[2\theta(x)-\Theta\right]}}
               {k(x)\sin{\Theta}}
        = g\s(x,\e) + g\osc(x,\e)\ .
\een
Thus,
\ben
\n\sc(x) = \oint_{C(\mu\sc)} \frac{d\e}{2\pi i}\, g\sc(x,\e) = \n\s(x) + \n\osc(x)\,,
\een
where $C(\mu\sc)$ is a contour enclosing all poles of
occupied states determined by $\mu\sc$.

First we evaluate the density coming from the smooth term.
In the limit $L\to\infty$ this is dominated by $\exp[-i\Theta]$, 
simplifying the integral to
\ben
\n\s(x) = -\frac{1}{2\pi}\oint_{C(\mu\sc)}\frac{\d\e}{k(x,\e)}\,,
\een
which, evaluated on the real axis, yields: 
\ben
\n\s(x) = \frac{1}{\pi}\int_{v(x)}^{\mu\sc}\frac{d\e}{k(x,\e)}
        = \frac{k\F(x)}{\pi}\ .
\een
\begin{figure}[htb]
\begin{center}
\includegraphics[angle=0,width=5cm]{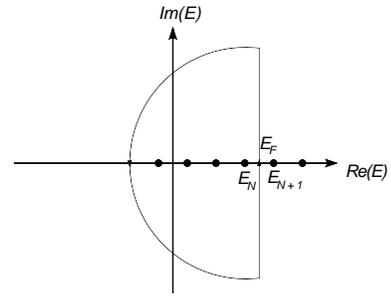}
\end{center}
\caption{Contour of integration in the complex $\e$-plane.}
\label{f:contour}
\end{figure}

Then, we evaluate the oscillating term of Eq.~(\ref{gf}). 
We pick a contour $C(\mu\sc)$ as shown in Fig.~\ref{f:contour}, 
i.e., a vertical line along $\e=\mu\sc+i\zeta$ 
connected to a semicircle, which encloses all poles of 
$N$ lower-lying energy-eigenvalues $\e_{N},\dots, \e_{1}$.
In the classical continuum limit  
$\mu\sc >> \zeta$, allowing us to expand all quantities in the integrand 
in powers of the imaginary part of the energy, $\zeta$:
\bea
\frac{1}{k(\mu\sc+i\zeta,x)} &=& \frac{1}{k\F(x)}\left(1-\frac{i\zeta}{k\F^2(x)}+\dots\right)\,,\\
\theta(\mu\sc+i\zeta,x) &=& \theta\F(x)+i\zeta\,\tau\F(x)+\dots\ .
\eea
Keeping terms up to first order in $\zeta$, 
employing the semiclassical quantization condition for the given boundary conditions
in Eq.~(\ref{wkbqc}) with $j=N+1/2$, and 
substituting $u=4T\F\zeta$,  
we obtain the result in Eq.~(\ref{nsc}).
Note that the additional term of $1/2$ in the quantization condition
relative to Ref.~\onlinecite{KSb65} is due to the Mazlov index for a
hard wall being $0$, rather than $1/4$ at a real turning point.

Next, we provide some details of the lengthy derivation
of the kinetic energy density of non-interacting, same-spin fermions:
\ben
t\sc(x) = \oint_{C(\mu\sc)}\frac{d\e}{2\pi i}\,[\e-v(x)]g\sc(x,\e)
        = t\s(x) + t\osc(x)\ .
\een
In analogy to the derivation of the semiclassical density we first evaluate 
the smooth term yielding
\ben
t\s^{(1)}(x) =\frac{k\F^3(x)}{6\pi}\ .
\een
The subdominant piece of smooth term gives another contribution:
\ben
t\s^{(2)}(x) = 
\oint_{C(\mu\sc)} \frac{d\e}{2\pi i}\frac{k(x,\e)\exp{i\Theta(\e)}}{4\sin{\Theta(\e)}}
\een
is evaluated on the contour $C(\mu\sc)$ as shown in Fig.~\ref{f:contour}.
Hence, all quantities in the integrand are expanded in powers of $\zeta$.
In particular, we define  $s=-2i\Gamma(\zeta)$,
where $\Gamma(\zeta)=\int_0^L dx\, [\sqrt{2[\mu\sc(x)+i\zeta]}-\sqrt{2\mu\sc(x)}]$,
express the $\zeta$-expansion of all quantities in terms of $s(\zeta)$,
and truncate its expansion after the quadratic term.
Note that we approximate terms like $[\int_0^L dx/k\F^3(x)]/T\F$ by $1/k\F^2(x)$.
This amounts to the same as neglecting the derivatives of $\kappa_j$
Then we obtain
\ben
t\s^{(2)}(x) = \frac{1}{2\pi k\F(x)T\F^2}
\int_0^{\infty}d s\ \frac{sz}{z-e^s}\,,
\een
where $z=\exp{[2i\Theta\F]}$, which yields 
\ben
t\s^{(2)}(x) = -\frac{\pi}{24k\F(x)T^2\F}\,,
\een
where the total contribution of the smooth term
is the sum of $t\s^{(1)}$ and $t\s^{(2)}$, agreement with Eq.~(\ref{ts}).
Then, we evaluate the oscillating term, which
is integrated also along the contour $C(\mu\sc)$ in Fig.~\ref{f:contour}.

We write the cosine of the oscillating piece as a weighted sum of exponential functions
and demonstrate the derivation for the term that has the positive sign 
in the exponential function. We call this term $t\osc^{(1)}$.

As before we expand all quantities in powers of $\zeta$.
Here, we define $q=-4i\Gamma(\zeta)$, express the $\zeta$-expansion
of all quantities in the integrand by $q(\zeta)$, and truncate its expansion
after the quadratic term. 
Then we integrate by aid of the polygamma functions of order $n$,\cite{AS72} defined as
$\psi^{(n)}(l)=(-1)^{n+1}\int_0^\infty dq\,q^{n}\exp{(-lq)}/[1-\exp{(-q)}]$, yielding
\bea\label{tpg}
&~~& t\osc^{(1)}(x) = -\frac{k\F(x)}{8\pi T\F}
\left[\psi\left(\frac{y+1}{2}\right)-\psi\left(\frac{y}{2}\right)\right]\sin{2\theta\F(x)}\nonumber\\
&~~&-\frac{1}{16\pi k\F(x) T\F^2}
\left[\psi^{(1)}\left(\frac{y}{2}\right)-\psi^{(1)}\left(\frac{y+1}{2}\right)\right] \cos{2\theta\F(x)}\nonumber\\
&~~&-\frac{1}{128\pi T\F^3 k\F^3(x)}
\left[\psi^{(2)}\left(\frac{y+1}{2}\right)-\psi^{(2)}\left(\frac{y}{2}\right)\right] 
\sin{2\theta\F(x)}\ .
\eea
Similarly, the other term $t\osc^{(2)}$ integrates to the result in Eq.~(\ref{tpg}) 
with $y\to -y+1$, where $y=\alpha\F(x)/\pi$. The particular combination of the polygamma 
functions in $t\osc^{(1)}$ and $t\osc^{(2)}$ yields
\bea
&~~& t\osc(x) = -\frac{k\F(x)\sin 2\theta\F(x)}{4 T\F\sin\alpha\F(x)}
-\frac{\pi\,\cos\alpha\F(x)\,\cos{2\theta\F(x)}}{4 k\F(x) T\F^2\sin^2\alpha\F(x)}\nonumber\\
&~~&-\frac{\pi^2\sin{2\theta\F(x)}}{8 T\F^3 k\F^3(x)\sin\alpha\F(x)}
\left(\half-\frac{1}{\sin^2\alpha\F(x)}\right) \ .
\eea
Finally, the sum of the smooth and oscillating pieces yield the result of Eq.~(\ref{tsc}).
\label{page:end}

\end{document}